# PHASE DIAGRAM FOR ANDERSON DISORDER: BEYOND SINGLE-PARAMETER SCALING

**Nigel Goldenfeld**,

Department of Physics,

University of Illinois at Urbana-Champaign, 1110 West Green Street,

Urbana, IL 61801, USA, and Department of Applied Mathematics and

Theoretical Physics, Wilberforce Road, Cambridge CB3 0WA, UK

and

**Roger Haydock**,

Department of Physics and Materials Science Institute,

1274 University of Oregon,

OR 97403-1274, USA.

ABSTRACT

The Anderson model for independent electrons in a disordered potential is transformed analytically and exactly to a basis of random extended states leading to a variant of augmented space. In addition to the widely-accepted phase diagrams in all physical dimensions, a plethora of additional, weaker Anderson transitions are found, characterized by the long-distance behavior of states. Critical disorders are found for Anderson transitions at which the asymptotically dominant sector of augmented space changes for all states at the same disorder. At fixed disorder, critical energies are also found at which the localization properties of states are singular. Under the approximation of single-parameter scaling, this phase diagram reduces to the widely-accepted one in 1, 2 and 3 dimensions. In two dimensions, in addition to the Anderson transition at infinitesimal disorder, there is a transition between two localized states, characterized by a change in the nature of wave function decay.

## 1. The Anderson Model

One of the fundamental questions about the electronic structure of a material is whether it is metallic or insulating, a distinction which reduces to whether the electronic states carry currents (time-reversal doublets), or whether they cannot carry currents (time-reversal singlets). The nature of electronic states depends on interactions between electrons and ions, the structure of the material, as well as on the interactions between electrons and other excitations. While the states of independent electrons in crystalline structures are well understood, the effects of structural disorder and interactions between electrons continue to be studied. Of these two, structural disorder seems the simpler, although many aspects of non-interacting electronic states in two-dimensional and three-dimensional disordered systems remain controversial. This paper addresses the breakdown, with increasing structural disorder, of the metallic state in which the quasi-electrons near the Fermi level move independently.

The Anderson model [1] is a minimal Hamiltonian for independent electrons in a disordered potential. The hopping part of the model is periodic with a single tight-binding orbital on each site of a lattice, which is taken here to be hypercubic of dimension D (a chain, square, or cubic lattice in D=1, 2, or 3, respectively), and with the same hopping matrix-element between each pair of nearest-neighbor orbitals.  The disordered structure produces a disordered electronic potential which is included in the model as disorder in the energies of the orbitals; they are taken independently from some distribution, usually a top hat.  The Anderson model neglects interactions between electrons, so it only applies to the quasi- electrons close to the Fermi level in a metal, where interactions become arbitrarily small.  Hence, this model describes the breakdown of a metallic state with increasing disorder.  In terms of operators $\{\phi_\alpha\}$ which annihilate electrons in the orbitals located on the lattice of sites $\{\mathbf{R}_\alpha\}$, the Anderson Hamiltonian is,

$$H = \sum_\alpha \varepsilon_\alpha \phi_\alpha^+ \phi_\alpha + h \sum_{\beta \in \alpha} \phi_\alpha^+ \phi_\beta, \qquad (1)$$

where $\varepsilon_\alpha$ is the energy of the orbital at $\mathbf{R}_\alpha$, taken independently for each orbital from the distribution $\rho(\varepsilon)$, $h$ is the hopping matrix-element, and the sums are over sites $\alpha$ and sites $\beta$ which are nearest-neighbors of $\alpha$.

Despite its simplicity, it is difficult to say much about the stationary states of the Anderson Hamiltonian, other than in the small or large disorder limits where the width $W$ of the distribution of site-energies is respectively much smaller or much larger than $h$. The problem is that the random energies make the orbitals inequivalent, and there are an infinite number of them differing by arbitrarily small energies. The consequence of this infinite quasi-degeneracy is that perturbation methods do not converge because energy denominators are too small. Moreover, the stationary states of finite subsystems do not converge with increasing size, because each state in a finite subsystem hybridizes strongly with the infinite number of states outside the subsystem which are arbitrarily close in energy.

For intermediate disorders, only quantities such as densities of states and related Green functions can be calculated and this must be done non-perturbatively. Scaling [2] is widely used to interpolate between the small and large disorder limits; and numerical approaches range from numerical scaling [3,4] to sampling the densities of states in various ways such as level distributions [5]. Numerical results are typically noisy due to slow convergence with the number of samples of orbital-energies, but despite the noise there seem to be significant disagreements between numerical methods, especially in two dimensions[6-13]. It is difficult to see how these issues can be resolved by advances in computers or computational methods, so analytic results are needed. In addition to single-parameter scaling [2] and apart from the work presented in this paper, there are two other analytical approaches to this problem: the first is the construction of effective field theories [14] leading to the non-linear sigma model, and the second is infinite-order perturbation theory [15] leading to diffusion poles in the electronic Green function.

In parallel with the efforts described above, there has been rigorous mathematical work [16-23] to establish the basic properties of the Anderson model. While the extended properties of states in ordered systems are well understood, one of the challenges to the rigorous approach has been to show that sufficient disorder makes all states exponentially localized [16]. One of the surprising properties of the Anderson model is that there is no feature in the density of states (averaged density of states, but *not* projected densities of states) [24] at energies separating extended from localized states. This has been shown rigorously for a number of examples [17, 18]. Although the work described in this paper characterizes the metallic phase as having broken time-reversal symmetry, the mathematical approach emphasizes the classification of spectra into point, absolutely continuous, and singular-continuous [17, 19].

The purpose of this article is to present an analytic approach to the Anderson Hamiltonian that does not depend on any scaling hypothesis [2], is non- perturbative [15], and does not proceed from the direct assumption of a Gaussian distribution of disorder potentials [14], and the concomitant reliance on field-theoretic methods. The problem with a scaling hypothesis is just that it is a hypothesis, and the problems with perturbation theory are mentioned above. Our concern with the Gaussian distribution is that the spectral properties of bounded operators can differ qualitatively from the spectral properties of unbounded operators. The Anderson Hamiltonian has the former quality, and this should be taken into account in calculating its properties. The field-theoretic approach works directly from a Gaussian distribution of disorder potentials [14], unbounded and not obviously valid for this system. Our approach shares something in common with the conventional field-theoretic approach, namely the strategy of removing the disorder so as to leave a pure, effective Hamiltonian in which the electronic degrees of freedom are no longer independent but experience an interaction from integrating out the disorder. However, since we purposefully reject the Gaussian distribution at the outset, our analysis uses a different methodology. In particular, it avoids the replica limit and the introduction of supersymmetry.

The first analytic method used in this work is projection, a transformation of the model in which only those states which couple to some particular state are retained. As is shown below, the Anderson model can be projected exactly and analytically onto extended states, with the weak law of large numbers eliminating disorder from the matrix-elements of the transformed model, because each one depends on an infinite number of infinitesimal contributions from independent orbital-energies with well behaved distributions. Disorder remains in the transformed model, but is represented in the basis set of extended states by expansion coefficients for the various site-orbitals which depend on the random site-energies. This transformed model is a variant of the augmented space representation[25,26] for the average electronic structure of random alloys; however the augmented space generated here is based on extended states rather than localized orbitals as in its application to random alloys. Recently, calculations using averaged quantities[27] in the spirit of the original derivation of augmented space [26] have produced phase diagrams intermediate between the results presented here and those of scaling theory.

For a physical picture of the Anderson model in augmented space, think of the extended states in the projection as sites on a new lattice. Taking this approach, the action of the hopping term in the Hamiltonian, the second term in Eq. 1, can be expressed as the sum of translations of the extended states by each of the nearest-neighbor displacements, hopping on the new lattice. The random potential, the first term in Eq. 1, multiplies the component of the extended state on each site by the value of the random potential at that site, producing a new extended state which is unrelated to any translation of the old one. In contrast to hopping, think of this new extended state as an internal change in the site on the new lattice, a kind of 'spin' on the new lattice. The resulting picture of augmented space is an electron hopping on a lattice of spins with an interaction between the electron and the spin on the site it occupies. The mathematical formulation of this transformation is presented in the next Section.

One important difference between the Anderson model in augmented space and position space is that while the number of sites in the Anderson model is countable, the number of extended states is uncountable. As a consequence of the uncountable dimension of this state-space, the evolution of the system is non-ergodic in the sense that, starting from a single extended state, the system can never explore more than an infinitesimal fraction of the extended states. As parameters in the model change, the system's evolution can shift from one subspace of extended states to another, producing one of the two kinds of phase transition in the model. Lest the reader be concerned that somehow the Anderson model has grown during projection, it is important to point out that the extended states are limits of combinations of site-orbitals, so the projected model is no larger than the original model including the limiting states. Indeed, it is just the limiting properties of states which determine whether they are metallic or insulating, and these limiting properties are explicit in projection.

In previous work[13], augmented space was used to obtain a variational expression for the edges of the band of states which are exponentially localized in augmented space. These band edges agree with perturbative results for the mobility edges at small disorder in three dimensions [10] and with localization edges in one and two dimensions[10]; and at intermediate disorder these band edges agree with numerical results obtained from the recursion method[4,12,13].

The second analytic method used here is path-counting in order to determine which parts of augmented space dominate the system's evolution. One of the puzzles of previous results[13] is that the width of the band of states, localized in augmented space, increases monotonically with disorder, even at disorders much larger than needed to localize all states in position space. In the previous paper it was argued that in low dimension or at large disorders where all states are localized in position space, these band edges are singularities in the localization length of exponentially localized states, and don't separate extended states from localized states.

Numerical results from the recursion method [13] suggested what is shown in this work, that there are in fact two distinct kinds of phase boundary in the Anderson model, both continuous in the sense that the energy of states (free energy at zero temperature) varies continuously with $W$ and $h$. The first kind of transition is characterized by a singular change in the sector of augmented space which dominates the asymptotics of the states as $W$ changes for fixed $h$. It includes the Anderson transitions at zero disorder in one and two dimensions as well as well as the Anderson transition at non-zero disorder in three dimensions, where the entire band of states changes its localization properties at a critical value of disorder. These transitions of the Anderson kind also include transitions from power-law to exponential, from extended to power-law, and possibly many other subtler transitions. It is the characterization of this first kind of transition which is the main result of this work. The second kind of transition is characterized as a singular change in localization properties as the energy varies for fixed disorder. It includes the mobility edges in three dimensions, but also includes localization transitions from power-law to exponentially localized states in two and three dimensions, as well as singular changes in the exponential localization length.

The paper is organized into seven further Sections. In Sec. 2 a new derivation of the transformation of the Anderson model to augmented space is presented. In the next Sec. asymptotic properties of the states in augmented space are related to the different phases of the Anderson model. The asymmetric Cayley tree is solved in Sec. 4 as a simplified version of the Anderson model in augmented space and this leads to an approximate phase diagram for the Anderson model. In Sec. 5, analytic expressions are obtained for critical disorders of the full Anderson model transformed into augmented space. The approximation of single-parameter scaling is applied to the phase diagram resulting from this work, and it is shown that this reproduces the usual scaling phase diagram in Sec. 6. In Sec. 7, the results of this work are compared with field-theoretic and perturbative approaches, and in the final section, the conclusions of this work are summarized.

## 2. The Anderson Model in Augmented Space

There are few analytic results for the Anderson model, especially away from the limits where the width of the distribution of site-energies is either very large or very small compared to $h$. The transformation to augmented space is analytic and exact, and applies for all disorders and all distributions of site-energies which are well-behaved, in the sense that their moments determine the distribution. The idea behind this transformation is projection of the model onto a basis $\{\Phi_s\}$ of distorted waves which are taken to be either states, or the operators which annihilate those states, depending on what is convenient in the context of the discussion. The distorted waves are constructed from polynomials $\{p_n(\varepsilon)\}$ which are orthonormal[28] with respect to the distribution $\rho(\varepsilon)$ of the site-energies in the sense that $\int p_n(\varepsilon) p_m(\varepsilon) \rho(\varepsilon) d\varepsilon$ is $\delta_{n,m}$. The transformation begins with $\Phi_0$, an extended state which has coefficient one for every site-orbital in the Anderson model. The general element of the new basis is

$$\Phi_s = \sum_\alpha \{\prod_\beta p_{s(\beta)}(\varepsilon_{\alpha+\beta})\} \phi_\alpha, \qquad (2)$$

where $s$ is a vector whose components $\{s(\beta)\}$ are the degrees of orthogonal polynomials $\{p_{s(\beta)}(\varepsilon)\}$, one for each site $\beta$, and $\varepsilon_{\alpha+\beta}$ is the energy of the orbital at $\mathbf{R}_\alpha + \mathbf{R}_\beta$. These states are extended over the entire system, but the coefficients of individual site-orbitals in these states depend on the disordered potential.

The most important property of the $\{\Phi_s\}$ is that they are orthonormal to one another, as is shown by the following argument. For extended states such as the $\{\Phi_s\}$, the inner product must be renormalized from the sum of products of corresponding components, to the average over all sites of the products of corresponding components,

$$(\Phi_s, \Phi_{s'}) = \left\langle \prod_\beta p_{s(\beta)}(\varepsilon_{\alpha+\beta})\, p_{s'(\beta)}(\varepsilon_{\alpha+\beta}) \right\rangle_\alpha \qquad (3)$$

In the above product, the distribution over $\alpha$ of $p_{s(\beta)}(\varepsilon_{\alpha+\beta})\, p_{s'(\beta)}(\varepsilon_{\alpha+\beta})$ is independent for each choice of $\beta$ because the energies of different sites are independent of one another. As a result, the average of this product over $\alpha$ is the product for different $\beta$ of the averages over $\alpha$ giving,

$$(\Phi_s, \Phi_{s'}) = \left\langle \prod_\beta p_{s(\beta)}(\varepsilon_{\alpha+\beta})\, p_{s'(\beta)}(\varepsilon_{\alpha+\beta}) \right\rangle_\alpha = \prod_\beta \left\langle p_{s(\beta)}(\varepsilon_{\alpha+\beta})\, p_{s'(\beta)}(\varepsilon_{\alpha+\beta}) \right\rangle_\alpha. \qquad (4)$$

From the weak law of large numbers, the averages, $\langle p_{s(\beta)}(\varepsilon_{\alpha+\beta})\, p_{s'(\beta)}(\varepsilon_{\alpha+\beta}) \rangle_\alpha$, over an infinite number of sites are simply the orthonormality relations for the polynomials: $(\Phi_s, \Phi_{s'})$ is just $\delta_{s,s'}$, that is zero unless $s$ and $s'$ are identical component by component.

The second important property of the $\{\Phi_s\}$ is that as a basis for the Anderson Hamiltonian, they produce a matrix which is sparse in the sense that each of the $\{\Phi_s\}$ has non-zero matrix elements with only a few others, as can be seen from the following argument. In additional to being orthonormal, the polynomials $\{p_n(\varepsilon)\}$ satisfy a three term recurrence relation[28], $\varepsilon p_n(\varepsilon)=b_{n+1}p_{n+1}(\varepsilon)+a_np_n(\varepsilon)+b_{n-1}p_{n-1}(\varepsilon)$, where the parameters $\{a_n\}$ and $\{b_n\}$ depend on the particular choice of distribution for the site-energies. So, when the first term of the Hamiltonian in Eq. 1 is applied to one of the $\{\Phi_s\}$ from Eq. 2, it multiplies each component for each site by the site-energy, and the recurrence for the polynomials relates this product to a sum of polynomials. As a result, the disordered potential changes the first component of the spin vector **s** as can be seen in the first three terms of the transformed Hamiltonian in Eq. 5. The second term of the Anderson Hamiltonian translates components of the $\{\Phi_s\}$ to each of their nearest-neighbor site, which is the same as translating the components of the spin vector **s** to their nearest neighbors, giving the fourth term in the transformed Hamiltonian,

$$H = \sum_s \{b_{n+1} \Phi_{s+1}^+ \Phi_s + a_n \Phi_s^+ \Phi_s + b_n \Phi_{s-1} \Phi_s + h \sum_\delta \Phi_{s'}^+ \Phi_s\}, \quad (5)$$

where $\delta$ is a nearest neighbor displacement, $s'(\beta+\delta) = s(\beta)$, $\{a_n\}$ and $\{b_n\}$ are the coefficients in the recurrence for the orthogonal polynomials, and **1** is the vector with unit component for the site at the origin and zero for all other components.

The localization of states is determined by their asymptotic properties in augmented space. For example, states whose contributions from the $\{\Phi_s\}$ decrease exponentially with the number of hops from $\Phi_0$ to $\Phi_s$ are dominated by just a few of the $\Phi_s$ near $\Phi_0$ so, like these $\Phi_s$, they are extended in position space. At the opposite extreme, states whose contributions from the $\{\Phi_s\}$ are independent of the number of hops from $\{\Phi_s\}$ are superpositions of many of the $\{\Phi_s\}$, and so cancel at most sites, because their components are random and independent, producing states which are localized in position-space.

In this asymptotic region, the hopping matrix-elements of H are still $h$ because the lattice is periodic, and from the theory of orthogonal polynomials [28], the asymptotic matrix-elements for the potential are determined by the edges of the distribution of site-energies (assumed continuous), which may be taken to be $\pm W/2$ without loss of generality. From this, the asymptotic value of the $\{b_n\}$ is $W/4$, and the asymptotic value of the $\{a_n\}$ is zero; whereas the $\{b_n\}$ diverge for a Gaussian distribution, indicated the important difference between bounded and unbounded distributions. Consider one of the basis states $\Phi_s$ in this asymptotic region of H, where almost all the sites near the origin have non-zero spins. First, $\Phi_s$ is coupled by *h* to 2D (a hypercubic lattice in D-dimensions) states where the spins have been shifted by a nearest-neighbor translation. Second, $\Phi_s$ is coupled by *W*/4 to two states for which the spin at the origin differs by ±1. This finite coordination of all states is one of the properties which make the Anderson model simple in augmented space.

The next important property of H is the structure of closed paths, sequences of matrix-elements which form loops. First there are the shortest closed paths, from $\Phi_s$ to one of its 2(D+1) neighbors and back, which do not even count as loops because there are no intermediate hops. The smallest loops are those associated with the original lattice, four *h*-hops around a square in two and higher dimensions, and the various larger loops for hypercubic lattices with D>1. The smallest loops which include both *h*-hops and *W*/4-hops are of length eight, alternating four *h*-hops and four *W*/4-hops. These loops begins with a change of the spin on **0**, a nearest-neighbor translation of all spins, another change of the spin on **0**, the inverse of the nearest-neighbor translation, the inverse of the first spin change, the first nearest-neighbor translation, the inverse of the second spin change, and finally the inverse of the nearest neighbor translation. A schematic of this structure is shown in Fig. 1. These and larger loops may be understood in terms of changing the configuration of spins by changing individual spins in different orders. Starting with one configuration we can change one of the spins by translating that spin to the origin, and then change another spin in the configuration by translating that one to the origin, and so on, until the desired final configuration is reached. Each pair of paths from the initial to final spin-configurations makes up a loop, provided there is not some intermediate configuration in common.

## 3. Phase Transitions in Augmented Space

An Anderson model is defined by its lattice, which is taken here to be hypercubic in D-dimensions, and one, dimensionless parameter $W/h$, the ratio of the width $W$ of the distribution of orbital-energies to the nearest-neighbor hopping matrix-element $h$. Models with the same lattice and $W/h$ have similar states. Hopping to more distant neighbors can be added, but there is no evidence this changes the model qualitatively unless the hopping is of infinite range. The states of the Anderson model have another dimensionless parameter $E/h$, the ratio of the energy of the state to $h$. A pair of states from different

Anderson models with the same lattice and *W/h* are almost certain to be similar if they have the same *E/h*.

In order for the model to exhibit different phases for different values of *E/h* and *W/h*, it must have states which differ qualitatively from one another. Finite combinations of site-orbitals are qualitatively the same, so the presence of multiple phases requires infinite numbers of site-orbitals, and it is the asymptotic behavior of states, i.e. their localization properties, which distinguishes the phases in this model. One goal of work on the Anderson model is to calculate the phase diagrams for different lattices in terms of *W/h* and *E/h*. This has proved difficult for reasons discussed above, and because it is always difficult to determine asymptotic properties of discrete equations such as the Schrödinger equation for the Anderson model.

Since the Anderson model in augmented space is a projection of the original model, the asymptotic properties of states in augmented space are related to those in position-space. The trend in this relationship is clear from the localization properties of combinations of the distorted waves which are the basis in augmented space, introduced in Sec. 2. These localization properties, localized to extended, invert in going from augmented space to position-space, but the boundaries between different phases occur at the same values of disorder and energy.

Analyzing the Anderson model in augmented space makes it possible to separate the properties of states which depend on *W* from those which depend on *E*, as is explained in what follows. In position space, the site-orbitals are countable, but in augmented space, the spin configurations {$\Phi_s$} are not countable, as can be seen by interpreting each spin as a digit in a real number. The importance of this larger basis is that the evolution of a single state, say $\exp\{-iHt\}$ $\Phi_0$, is a superposition of the powers of H on $\Phi_0$, a space of countable dimension, and therefore an infinitesimal fraction of augmented space which has uncountable dimension. As a result, it is possible for $\Phi_0$ to evolve onto different subspaces of countable dimension depending only on *W* through H, and this makes possible phase transitions which depend only on *W*, not on *E*, the Anderson transitions. Within subspaces of countable dimension which support the evolution of $\Phi_0$, there can still be states with different localization properties, so there is a second qualitative distinction between states and hence a second kind of phase transition, the mobility transitions, of this model.

## 4. Exactly Solvable Models

As an example of a model in augmented space with phase transitions similar to the Anderson model, consider a Cayley tree with coordination four – each vertex on the tree is connected by edges to four other vertexes, and there are no loops in the graph. Take the vertexes of the tree to represent basis states in augmented space, and take the edges of the tree to represent non-zero matrix-elements of the Anderson Hamiltonian in augmented space. From each vertex, take two of the edges to have hopping matrix-elements $h$, assumed positive, and take two edges to have matrix-elements of the potential $W/4$. This is illustrated in Fig. 2. The asymmetric Cayley tree differs from the symmetric Cayley tree in having more than one kind of edge.

Locally (out to fourth neighbors), the asymmetric Cayley tree is identical to the Anderson model in augmented space far from $\Phi_0$ for an infinite chain of equally spaced sites, D=1, with a semi-elliptical distribution of orbital-energies of width $W$ ($a_n=0$, $b_n=W/4$). At fourth neighbors, changing spins in different orders leads to the same state for the Anderson model, but different states for the Cayley tree. Another difference between the two models is that all vertexes of the Cayley tree are equivalent, but for the Anderson model $\Phi_0$ is special.

When *h* exceeds *W*/4, it is clear that the system evolves preferentially along *h*-hops so that the asymptotics of states are dominated by *h*-hops, one phase of the system. When *W*/4 exceeds *h*, *W*/4-hopping is preferred, and the asymptotics of states are dominated by *W*/4-hops, the second phase of the system. When *h*=*W*/4, the tree is symmetric as is the propagation of the system, and the asymptotics of the states is critical. For the asymmetric Cayley tree all the states change character as the disorder *W* moves through its critical value of 4*h*, so this is an example of the Anderson type of phase transition in augmented space.

A mobility transition occurs as *E* varies for fixed *W*. The simplest example of this kind of transition occurs at the critical value of disorder $W_c$, where both kinds of edges have the same matrix-elements and the Cayley tree is symmetric. The energies of stationary states coupled to a particular basis state are those which contribute to the projected density of states for a single vertex of the tree. The simplest way to calculate this is with a hierarchy of equations for the projected resolvent (or Greenian),

$$R(E) = <\Xi_0 [E - H_C]^{-1} \Xi_0^+>, \qquad (6)$$

where $H_C$ is the Hamiltonian for the Cayley tree, $\Xi_0$ annihilates the state represented by one of the vertexes on the tree, and the angle brackets mean the vacuum expectation value.

The projected resolved R(E) can be calculated by relating it to other projected resolvents, noting that site $\Xi_0$ is coupled to four neighboring vertexes by the matrix-element $h$ (spin-hops and lattice-hops have the same matrix-elements in the symmetric case). Define a second projected resolvent,

$$G(E) = <\Xi_1[E - H_{C0}]^{-1}\Xi_1^+>, \quad (7)$$

for any of these neighboring vertexes (they are equivalent) with the Hamiltonian $H_{C0}$, which is $H_C$ with $\Xi_0$ removed. Some matrix algebra gives the first of the hierarchical equations,

$$R(E)^{-1} = E - 4\,h^2\,G(E). \quad (8)$$

The second equation comes from noting that $\Xi_1$ is coupled by $h$ to three vertexes other than $\Xi_0$, and the projected resolvents for these vertexes (excluding $\Xi_1$ from the Hamiltonian) are also G(E) because the Cayley tree is infinite with equivalent vertexes, leading to

$$G(E)^{-1} = E - 3h^2 G(E). \quad (9)$$

Equation 9 is quadratic in G(E), so it can be solved and the result substituted into Eq. 8 to give R(E). For real E, the imaginary part of R(E) is π times the projected density of states which comes out to be,

$$n(E) = (2/\pi)\sqrt{(12h^2 - E^2)/(16h^2 - E^2)}, \text{ for } -2\sqrt{3}h \le E \le 2\sqrt{3}h,$$
and zero otherwise. (10)

Note that the model has states with energies between -4h and +4h, the extremal states are marked by the zeros of the denominator in Eq. 10, but only those between -2√3h and +2√3h couple to single basis states. The states in the range ±2√3h decrease exponentially with distance (number of edges) steeply enough to have significant weight on the original vertex, while the states outside this range do not decrease with distance steeply enough for there to be any density of states on the original vertex. For example the state at $E=4h$ is 1 on every vertex of the tree, while the state at $2\sqrt{3}h$ has amplitude ½ on each of the nearest neighbors of the original vertex, $1/(2\sqrt{3})$, on each of the next neighbors, 1/6 on each third neighbor, and so forth decreasing by a factor of $1/\sqrt{3}$ on each succeeding shell after the first. As a result, there are two critical energies $E_c=\pm 2\sqrt{3}h$ where the projected density of states, and hence the localization properties of states, are singular for this model; and there are the two extremal energies of the model $E_L=\pm 4h$, the Lifshitz edges, at which the properties of states are also singular.

      Away from the critical disorder, there are similar critical energies, but the algebra gets more complicated. As before, the simplest approach is to set up hierarchical equations for projected resolvents. Take $R(E)$ to be the projected resolvent for one vertex of the tree. It is coupled by lattice-hops to two vertexes whose projected resolvents are taken to be $G(E)$, defined with a Hamiltonian excluding the original vertex making $G(E)$ different from $R(E)$. The original vertex is also coupled by spin-hops to two vertexes whose projected resolvents are taken to be $S(E)$, again defined by a Hamiltonian excluding the original vertex so $S(E)$ is different from $R(E)$ and $G(E)$. In terms of $G(E)$ and $S(E)$,

$$R(E)^{-1} = E - 2 h^2 G(E) - W^2 S(E)/ 8. \qquad (11)$$

Similar arguments lead to equations for G(E) and S(E),

$$G(E)^{-1} = E - h^2 G(E) - W^2 S(E)/ 8, \qquad (12)$$

$$S(E)^{-1} = E - 2 h^2 G(E) - W^2 S(E)/ 16. \qquad (13)$$

Equations 11-13 combine to give a quartic equation for R(E), which is where the algebra becomes complicated. Just as for the symmetric case there are two critical energies, which are the edges of the band of states sufficiently localized to contribute to the projected density of states on a single vertex. In addition there are two Lifshitz edges at $E_L=\pm(2h+W/2)$ which are the energies of the state taking value one on each vertex and the state alternating plus and minus one on neighboring vertexes. Figure 3 shows the phase diagram for the asymmetric Cayley tree in augmented space including both the Anderson transition for all energies at the critical disorder, and the mobility transitions at critical energies for various disorders. Since the states of the Anderson model in augmented space form a continuum in the energy-disorder plane of Fig. 3, the transitions between different phases are continuous.

## 5. Critical Disorders and Energies for the Anderson Model

For the Anderson model in augmented space, loops in the graph make the phase diagram much more difficult to calculate and much more complicated than the phase diagram for the asymmetric Cayley tree, which has no loops. One source of loops in the Anderson model is that spins can be changed in different orders to produce the same state, and another is that for D greater than one, the lattices also contain loops. In contrast, the asymmetric Cayley tree has no loops in the lattice, because it is one-dimensional, and changing the spins in different orders leads to different vertexes on the tree. When loops are present, determining asymptotics of states is more complicated than the comparison of products of matrix-elements which works for a tree.

In common with the asymmetric Cayley tree, the Anderson model in augmented space has two different kinds of matrix-elements: those for spin-hops and those for lattice-hops. This, together with the uncountable basis, makes possible multiple phases depending upon which directions in augmented space dominate the evolution of the system. At time $t$, the coefficient of $\Phi_s$ in a state which started as $\Phi_0$ at $t=0$, is

$$\psi_s(t) = (2\pi i)^{-1} \int <\Phi_0 | (E - H)^{-1} | \Phi_s> e^{-iEt} \, dE \qquad (14)$$

where the integral is around a contour which encloses the energies of all states. The leading term in the time-dependence of $\psi_s(t)$ is $\mu_s(-it)^N/N!$, where $N$ is the smallest power of H having a non-zero matrix-element $\mu_s$ between $\Phi_0$ and $\Phi_s$. In terms of the matrix elements of H, $<\Phi_0|H^N|\Phi_s>$ is the sum of products of matrix-elements of H along all the paths with the minimum number of hops $N$ from $\Phi_0$ and $\Phi_s$, the direct paths. Consequently, as **s** goes to infinity, $N$ goes to infinity, and the leading contribution of $\Phi_s$ to the state is proportional to $\mu_s$. Hence, the phase of the system is determined by the distant $\Phi_s$ with the largest $\mu_s$; that is the distant vertexes whose direct paths dominate, consistent with the result for the asymmetric Cayley tree.

In the case of the asymmetric Cayley tree, the asymptotic behavior of states is easy to determine from the above arguments, and it is shown here that they lead to a critical disorder, $W_c=4h$, which was derived in Sec. 4 by a different argument. Note that on a Cayley tree there is a unique direct path of $N$ hops from the origin to a given vertex. The moment $\mu_s$ is the product of matrix-elements, either $h$ or $W/4$, along the direct path. Hence the asymptotics of states on the asymmetric Cayley tree are dominated by the vertexes whose direct paths consist entirely of either $h$-hops or $W/4$-hops, whichever is larger, leading to an Anderson transition when they are equal.

Turning now to the Anderson model in augmented space, for $W=0$, $\Phi_0$ and other Bloch states are stationary solutions of the Hamiltonian, so there are no asymptotic tails in augmented space. However for $W$ infinitesimally greater than zero, the states acquire asymptotic tails, so there is a qualitative change in the states and hence a critical disorder $W_0=0$.

For $W$ greater than zero, but still small compared to $h$, lattice-hops must dominate the asymptotics of the states. The range of $W$ for which this dominance persists can be calculated using the arguments above. For hypercubic lattices in D-dimensions, the sites with the largest weight from direct paths of length Dn are located at $(\pm n, \pm n, \ldots \pm n)$ because this gives the maximum number of permutations of the lattice-hops in different directions. The contribution from these direct paths is $h^{Dn}(Dn)!/(n!)^D$. Using Stirling's approximation for the factorial, this becomes $(Dh)^{Dn}$, for n large, and is precisely the value which gives the exact width of the band of electronic states for the ordered lattice. Since the matrix-element for spin-hops is $W/4$, there is a transition at the critical disorder $W_1$ which leaves the contribution from direct paths unchanged when a lattice-hop is replaced by a spin-hop: $W_1=4Dh$.

It might seem that for $W$ greater than $W_1$, paths consisting only of spin-hops should dominate, but this is not the case, because of loops due to the equivalence of changing spins in different orders. Direct paths are generated by hops away from the origin, and those which alternate spin-hops with lattice-hops are especially numerous, because after each outward spin-hop, there are not just D, but 2D, outward lattice-hops. As a result, paths which contain mixtures of lattice-hops and spin-hops can outweigh the paths which are pure spin-hops up to the limit where $W/4=2Dh$. This then corresponds to a second critical disorder $W_2=8Dh$. Counting the number of direct paths of various lengths from $\Phi_0$ to $\Phi_s$ is a difficult combinatorial problem, so it is not possible to say where in the interval between $W_1$ and $W_2$ there are additional critical disorders. However, we can estimate bounds on critical disorders: $\Phi_s$ can be reached by direct paths which are mixtures of spin and lattice-hops, and which dominate the purely spin-hopping paths, for disorders less than $4\sqrt{2}Dh$ (5.657 D$h$) or even $4(3^{1/3})Dh$ (5.769 D$h$).

There are possibilities of additional phases in the Anderson model.

For example, in the case of the top hat distribution of site-energies, the matrix-element between $\Phi_0$ and $\Phi_1$ is $(W/4)/(\sqrt{3}/2)$ which is greater than $W/4$, enhancing contributions from paths which include spin-hops as well as lattice-hops. As a result, in this case there should be a transition at $W=4Dh\sqrt{3}/2$, but presumably this is very weak, and has not yet been detected numerically. Since other matrix elements for the spin-hops can be greater than their asymptotic value, additional transitions are possible corresponding to different patterns of spin and lattice-hopping.

In addition to this plethora of critical values of the disorder, the Anderson model has critical energies where the density of states is singular. These include van Hove singularities for ordered systems, but extend to singularities in localization lengths, in power-laws, and in other localization properties for disordered systems. The exact trajectories for mobility edges of the asymmetric Cayley tree in Sec. 4 can be used to approximate the mobility edges of hypercubic lattices in the Anderson model by equating ordered bandwidths: $h$ is replaced by $Dh$ in Eqs. 11-13. This is not a variational bound as was used in previous work [13], but could be further improved by taking into account variations in the number of paths to different sites on the hypercubic lattices. These critical energies together with the critical disorders are included in the phase diagram for the Anderson model in Fig. 4. Again, the states fill the energy-disorder plane and so the transitions are continuous.

It is useful to relate different phases in augmented space to their localization properties in position space. This can be done for the states at $W=0$ which are Bloch states and clearly extended in position-space. The states of models with $W>W_2$ consist of superpositions of random extended states in position space. The random extended states cancel on almost all sites, so in the extreme case these states are localized on a single site, where the random extended states happen to interfere constructively, with energy close to the random energy of that single site. The large disorder limit of the Anderson model in position space gives this same result [29]. For $D=1$, i.e. one dimension, and $W$ greater than zero, all states are exponentially localized, so as disorder increases, the exponential localization becomes stronger and the various phase boundaries correspond to singularities in the dependence of the localization length on disorder or energy. Analytic results for small but non-zero disorders show that in $D=2$ the states are power-law localized, and in $D=3$ the states are extended [10]. Beyond these results, all that can be shown analytically is that states become more localized with increasing disorder and that the nature of the localization depends on the dimension of the system.

## 6. The Single-Parameter Scaling Approximation

Numerical approaches to Anderson localization always require extrapolation from some finite system to an infinite limit. One well established method for doing this is finite-size scaling, which was applied by Pichard and Sarma [30, 31, 32] to the dependence of localization lengths on the width of long strips for two dimensions, and to the dependence of localization lengths on the side of long bars for three dimensions. This work showed that two dimensions is marginal for the existence of extended states, and that power-law localization is present for weak disorder in two dimensions, but the conclusion of this work in Ref. 32 is that exponential localization takes over at the longest length scales.

Single-parameter scaling makes the phenomenological assumption that the metallic or insulating nature of a material is determined by the variation of just a single parameter, the conductance $g$ across a hypercube (wire, square, or cube) with sides of length $L$. This approximation may be understood as building the wave functions of large hypercubes out of those of small hypercubes by matching only their amplitudes at the boundaries and neglecting the phase. In Ref. 2, the Authors go on to suppose that a smooth, non-decreasing, scaling function $\beta(g(L)) = d \ln g(L) / d \ln L$ interpolates between the limits of large $g$ in which $\beta(g(L))$ is Ohmic taking the value D-2 (where D is the dimension of the system) and small $g$ in which $\beta(g(L))$ is insulating and goes to Ln $g(L)$. If $\beta(g(L))$ is positive, the system scales to the metallic limit, and if $\beta(g(L))$ is negative, it scales to the insulating limit. For the critical value of the conductance separating metallic from insulating, $\beta(g(L))$ is zero.

The scaling function $\beta(g(L))$ can be calculated for different parts of the phase diagram in Fig. 4 using previous results from analytic and numerical recursion [10, 12, 13]. These results give the quantum mechanical transmittance for an electron starting on a single site to propagate either to infinity (analytic) or to a large distance, of order a thousand lattice constants, (numerical). The work of Landauer and Büttiker [33] relates this transmittance to a conductance from which Ohm's law gives a conductivity, taking into account the geometry of the numerical recursion, which is conduction from the boundary of a single site to the boundary of a large or infinite realization the model. This is simply the conduction from a small circle to a large concentric circle in two dimensions, or from a small sphere to a large concentric sphere in three dimensions. The conductance of the hypercube is independent of its size in two dimensions, but increases as the length $L$ of the edge in three dimensions.

For the one-dimensional Anderson model, recursion is trivial and non-zero disorder leads to a negative value of the scaling function producing the insulating limit, consistent with the scaling phase diagram in which there is an Anderson transition at zero disorder. For arbitrarily small disorder in two dimensions, both analytic and numerical recursion calculations [10, 12, 13] show that the transmittance away from a single site decreases as a negative power of the distance leading to $\beta(g(L)) < 0$. For the scaling function to take its critical value zero, the transmittance would have to decrease as $1/\mathrm{Ln}\, L$, slower than any power-law. For the two-dimensional Anderson model, this approximation makes all states insulating except those within the band at zero disorder, again consistent with the scaling phase diagram in which there is just an Anderson transition at zero disorder.

For three dimensions, a similar calculation shows that the scaling function takes its critical value zero for a transmittance which decreases as $1/L$ with distance $L$ from a single site - a critical power-law. Analytic recursion [10] for three dimensions shows that for small disorder the transmittance away from a single site goes to a non-zero constant for large $L$, scaling to metallic. The asymptotics of states in augmented space shows that there is an Anderson transition in three dimensions at $W=12h$, and numerical recursion [12,13] shows that this transition is to a power-law dependence of the transmittance on $L$.

From [12] the trajectory of this critical power-law can be roughly estimated, and is sketched as a thin line in Fig. 4. There is qualitative agreement between this trajectory and the most detailed phase diagram obtained from numerical scaling, Fig. 1 from Ref. 7. We now attempt a comparison between these numerical results and the present analytic results.

There is quantitative agreement between the present work and numerical scaling on the critical disorder for the Anderson transition in the following sense: In Fig. 4 of the present paper, the Anderson transition at $W=12h$, $D=3$, marks the disorder at which the slope of the scaling phase boundary is a minimum for positive energies; at larger disorders this phase boundary curves back to lower energies. In Fig. 1 of Ref. 7, this same minimum in the slope of the phase boundary also occurs at $12h$ to within the accuracy it can be estimated from that figure. There is also quantitative agreement between the disorder at which the phase boundary in Fig. 1 of Ref. 7 crosses zero energy, and the disorder at which the transmittance calculated in Ref. 12 decreases as $1/L$. However, this not a stringent test, because the resolution in disorder of the calculations in Ref. 12 is only $\pm 2$.

On the other hand, there is quantitative disagreement between this work and Ref. 7 in the placement of the mobility edges for $W=12h$, at $10.2h$ and about 7.9$h$, respectively. The energy resolution of the calculations in Refs. 12 and 13 is much better than the disorder resolution, and from [13] this critical energy lies between 8.5$h$ and 9.5$h$, which is inconsistent with the calculations presented here for the asymmetric Cayley tree, where we obtained the estimate of 10.2$h$. In Ref. [7], the estimate obtained also lies outside the range from Ref. [13]. For disorders less than 12$h$, our calculation of the phase boundary from the asymmetric Cayley tree is consistent with the lower and numerical bounds in Ref. [13]; however, the calculation of Ref. [7] is not consistent with either of those of Ref. [13]. Specifically, the phase boundaries of these various approaches are nested: the phase boundary in Ref. 7 lies inside the lower bound from Ref. 13 (a violation of the bound), which lies inside the numerical phase boundary from Ref. 13 (consistent with the bound), which lies inside the analytic phase boundary calculated from the asymmetric Cayley tree in Sec. 4 (also consistent with the bound).

In evaluating the above comparisons, it is important to keep in mind the inconsistency between the symmetry-breaking and scaling definitions of 'metallic' states. States with small power-law localization belong to time-reversal singlets and so are insulating by symmetry, however their conductivity according to [33] and Ohm's law scales to infinity on infinite length-scales. The reason for this inconsistency is that the metallic scaling limit depends on Ohm's law for which inelastic processes are implicit; however the Anderson model has no interactions and hence no inelastic processes.

    The main difference between the results of this work and those of scaling is that the projection onto augmented space reveals a much wider range of qualitative behaviors for the asymptotics of states.  Of particular importance is the time-reversal symmetry of states which is not addressed by scaling theories.  States belonging to time-reversal doublets are clearly metallic because they can carry currents, but scaling includes as metallic some states which are time-reversal singlets, namely those which are weakly power-law localized. While the asymptotics of some insulating states fall into simple classes such as exponential localization or power-law localization, augmented space reveals additional, more subtle, distinctions which have not yet been characterized. Compounding the complexity of the model are the singularities in the energy-dependence of states for a fixed disorder; sometimes qualitative changes such as between time-reversal singlets and doublets, and sometimes just singularities in a quantity such as the exponential localization length.

## 7. Comparison with Other Approaches

This work cannot be compared directly with field-theoretic approaches [14] because the field theories are defined on a continuum rather than a lattice. However, it is clear when the bounded top-hat distribution of site-energies is replaced by the unbounded Gaussian distribution in the transformation to augmented space, the matrix-elements for spin-hops diverge according to the recurrence relation for Hermite polynomials [28] which are orthogonal with respect to the Gaussian distribution. As a result, at sufficiently long times, spin-hops always dominate the evolution of states, and the model has only a single, strongly insulating phase for non-zero disorder. However, comparing with the usual interpretation of the field-theoretic results, there is qualitative agreement with the work in this paper for three dimensions, but qualitative disagreement with this work in two dimensions where the transition to exponential localization of all states occurs when $W=8h$.

Application of infinite-order perturbation theory to the Anderson model produces diffusion poles[15]. While electronic transport must certainly be diffusive if the electrons in the Anderson model are coupled to a heat bath, uncoupled, the states of the model have infinite lifetimes. In the metallic phase where the states belong to current-carrying, time-reversal doublets, these currents also have infinite lifetimes, which seems inconsistent with diffusion.

The results presented above agree with the basic requirements of the rigorous mathematical approach [16-23]: States are extended at zero or low disorder and exponentially localized at sufficiently strong disorder. The average density of states is smooth across mobility transitions, although this is not explicitly proved here. Turning to the spectral classification of states, it is clear that states belonging to time-reversal doublets have normalizations proportional to the volume of the system, and so are part of the absolutely continuous spectrum of the model. States belonging to the point spectrum have normalizations independent of the volume of the system, normalizable, and this spectrum is indeed dense. In addition, this work produces states whose normalizations are proportional to a power of the volume of the system greater than zero and less than one, so they belong to the singular-continuous spectrum whose existence has been hypothesized in the mathematical approach, but not demonstrated.

## 8. Conclusions

In this paper the critical assumptions of several approaches to Anderson localization have been tested by the transformation to augmented space which leads to exact values for the critical disorders of some Anderson transitions. These results agree with single-parameter scaling in that the Anderson transitions to non-conductive states occur at infinitesimal disorder in one and two dimensions, but not in three dimensions. The results disagree with single-parameter scaling in two dimensions in that there is an additional Anderson transition at non-zero disorder in two dimensions, between two non-conductive states which differ in the way that they decay at large distances. Furthermore, the phase diagram resulting from this work reduces to that of numerical scaling when the assumption of numerical scaling is combined with results from numerical recursion. What this work adds to the results of numerical scaling is the distinction between Anderson and mobility transitions, exact values for some of the Anderson transitions, and the existence of more Anderson transitions than was previously suspected.

In comparing this work with the results of field-theoretic methods, the general pictures agree: Anderson transitions between current-carrying and non-current-carrying states occur at zero disorder in one and two dimensions, but not in three dimensions. As in the case of scaling theory, the differences are in the additional Anderson transitions. One of the purposes of this work is to test the hypothetical equivalence of the bounded potential distributions in the Anderson model with the unbounded Gaussian distributions in the field theory. Transformation to augmented space of an unbounded distribution of potentials leads to domination of the disorder, making it very difficult to see how the states of the Anderson model can be other than exponentially localized for a Gaussian potential distribution. Because of this, the general agreement between this work and field-theoretic approaches is surprising - it may be that other approximations in the field-theoretic approach effectively cut off the tails of the Gaussians.

The representation of the Anderson model in augmented space incorporates both the states of the model and their limiting states which determine the phases of the model. Path counting and calculation of projected densities of states allow the phase boundaries to be located. That such a simple model as the Anderson model for quantum states in the presence of disorder can produce such a complicated phase diagram continues to surprise and delight.

**Acknowledgments**

We are grateful to David Khmelnitskii, Ben Simons, Dietrich Belitz and Eduardo Fradkin for helpful discussions. Support from the Richmond F. Snyder Fund and the University of Oregon Foundation is gratefully acknowledged by RH, as is the hospitality of the Condensed Matter Theory Group of the Cavendish Laboratory and Pembroke College Cambridge during part of this work. NG thanks the Department of Applied Mathematics and Theoretical Physics for its hospitality during the course of this work.

**Figure Captions**

Figure 1. Part of the local structure of the Anderson model in augmented space.

Figure 2. Part of the local structure of an asymmetric Cayley tree, approximating the Anderson model in augmented space.

Figure 3. The phase diagram in energy and disorder for the asymmetric Cayley tree, approximating the Anderson model.

Figure 4. The main phase boundaries for the hypercubic Anderson model in D dimensions, with a thicker line for the phase boundary for changes in time-reversal symmetry in D=3, and a thinner line for the contour of $1/L$ power-law states in D=3.

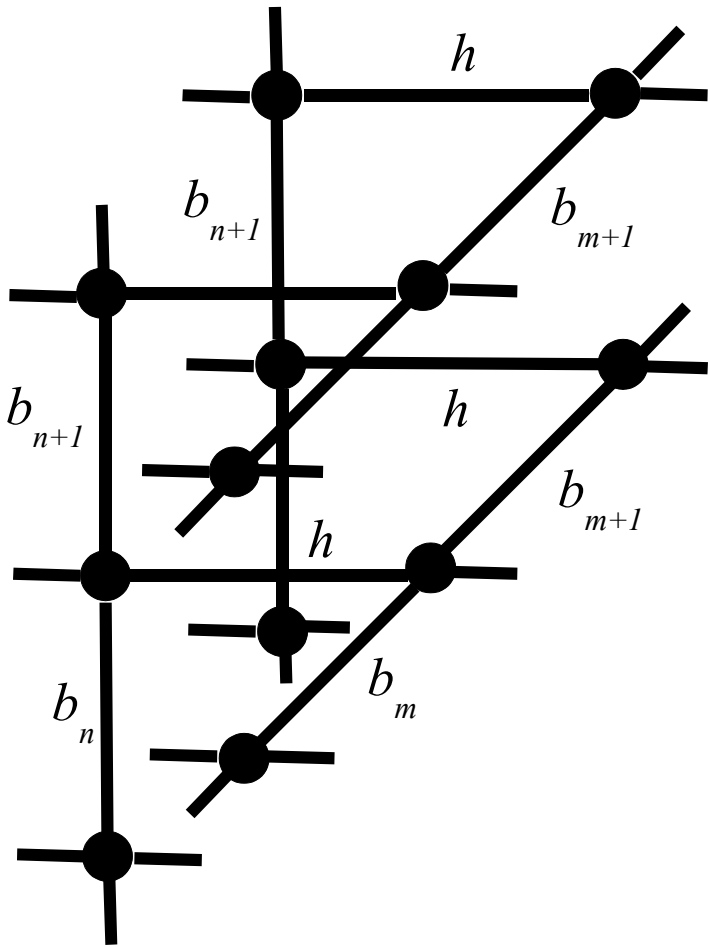

Figure 1

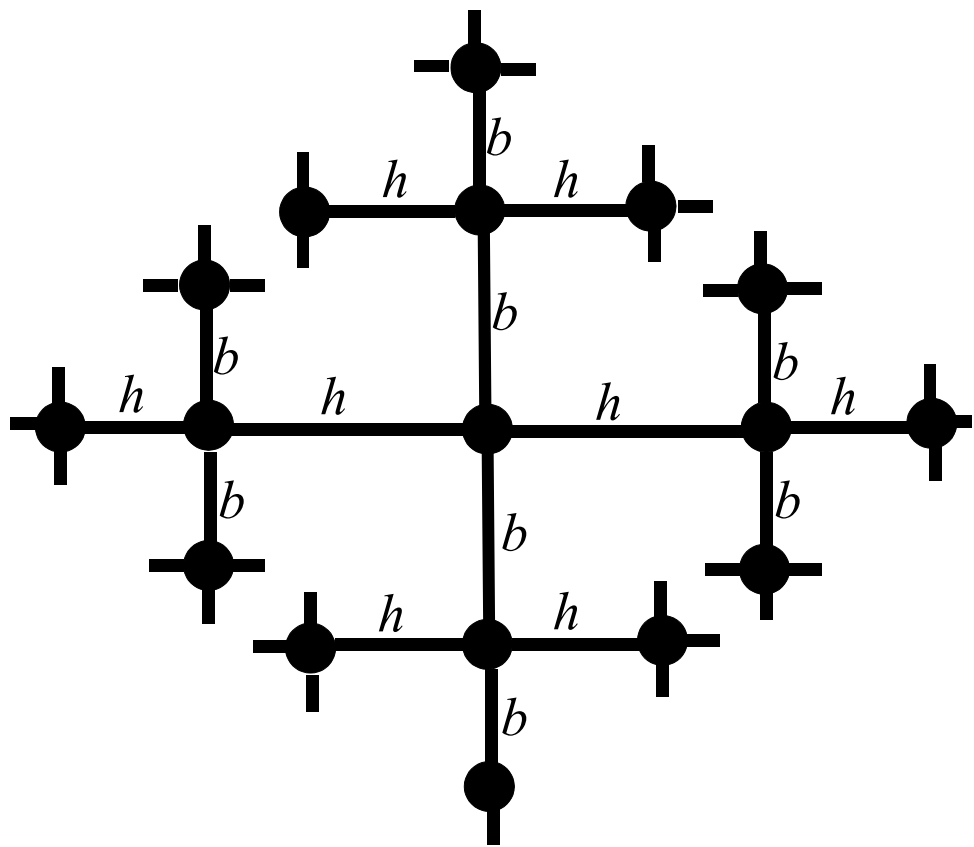

Figure 2

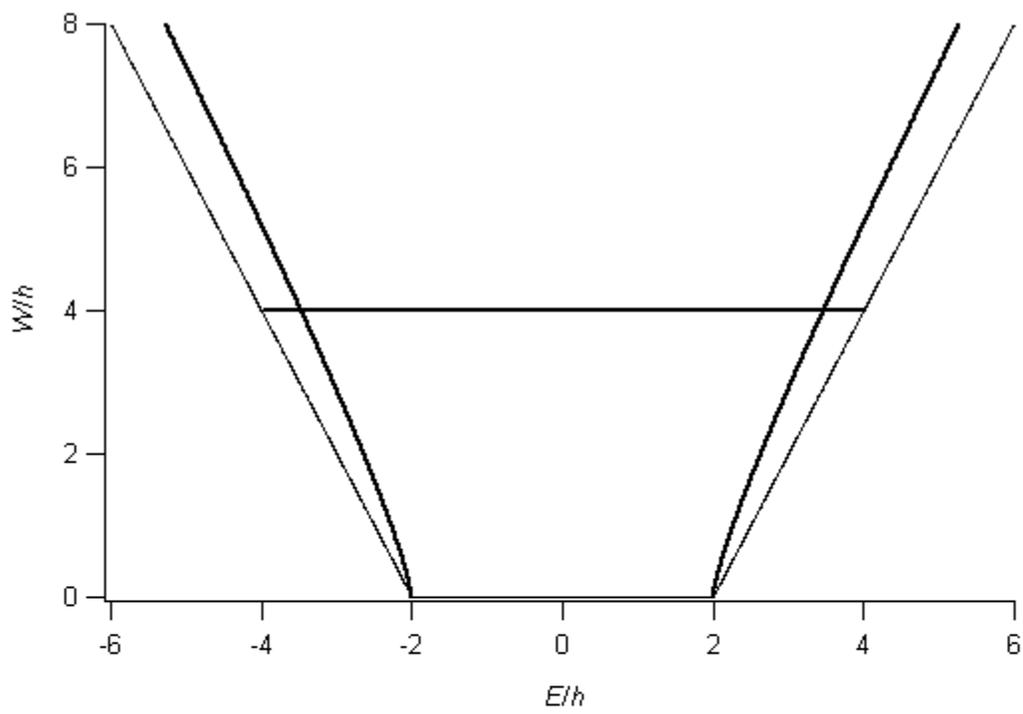

Figure 3

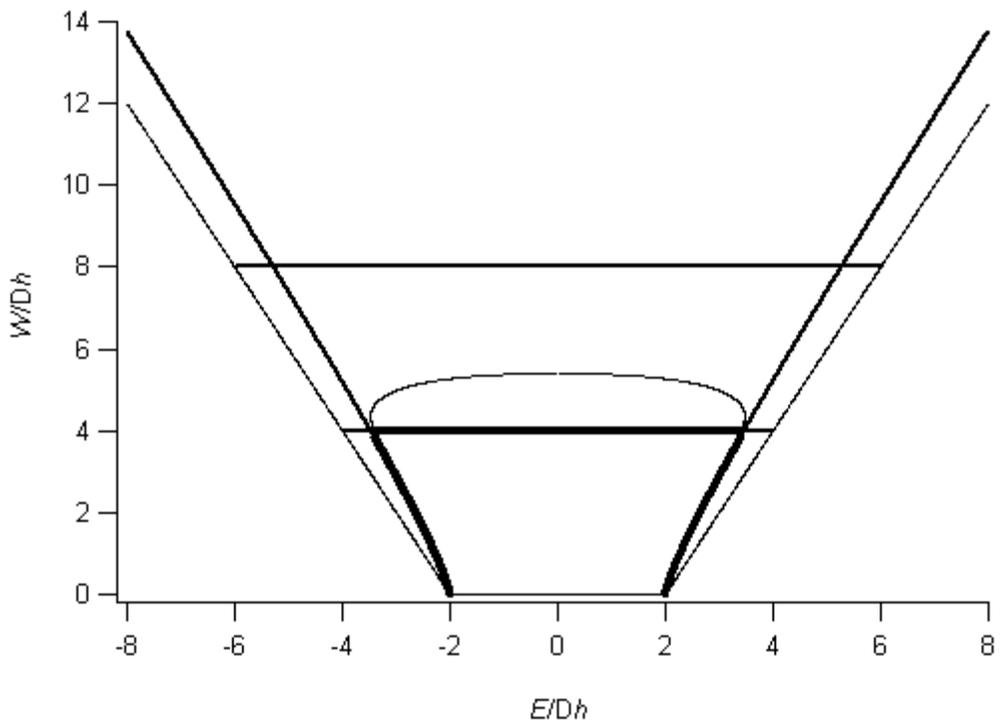

Figure 4